\documentclass[conference]{IEEEtran}

\usepackage[left=54pt, right=54pt, bottom=54pt, top=54pt]{geometry}

\usepackage{makecell}
\usepackage{pifont}   
\usepackage{adjustbox}
\usepackage{pgfplots}
\usepackage{cite}
\usepackage{amsmath,amssymb,amsfonts}
\usepackage{algorithmic}
\usepackage{graphicx}
\usepackage{textcomp}
\usepackage{array}
\usepackage{comment}
\usepackage[nolist]{acronym}
\usepackage[normalem]{ulem}
\usepackage{color,soul}
\usepackage{multirow}
\usepackage{multicol,multirow,colortbl}
\usepackage{enumitem}
\usepackage{url}
\usepackage{algorithm}
\usepackage[dvipsnames]{xcolor}
\usepackage{subcaption}

\UseRawInputEncoding

\newcommand{\bazzi}[1]{\textcolor{blue}{#1}}
\newcommand{\bazziremove}[1]{\bazzi{\sout{#1}}}

\newcommand{\lorenzo}[1]{\textcolor{purple}{#1}}

\begin{acronym} 
\acro{3GPP}{Third Generation Partnership Project}
\acro{5G}{fifth generation}
\acro{API}{application programming interface}
\acro{C-ITS}{cooperative-intelligent transport systems}
\acro{C-V2X}{cellular-vehicle-to-everything}
\acro{CAD}{computer-aided design}
\acro{CAM}{Cooperative Awareness Message}
\acro{CAV}{connected and autonomous vehicle}
\acro{CCAM}{cooperative connected and automated mobility}
\acro{CPM}{collective perception message}
\acro{DCC}{Decentralized Congestion Control}
\acro{DENM}{decentralized environmental notification message}
\acro{DSRC}{dedicated short range communication}
\acro{ETSI}{European Telecommunications Standards Institute}
\acro{FIFS}{first-in-first-scheduled}
\acro{FFT}{fast Fourier transform}
\acro{GNSS}{global navigation satellite system}
\acro{GPS}{global positioning system}
\acro{HAT}{hardware attached-on-top}
\acro{HMI}{human-machine interface}
\acro{ICW}{intersection collision warning}
\acro{IMU}{Inertial Measurement Unit}
\acro{IQ}{in-phase and in-quadrature}
\acro{ISAC}{integrated sensing and communication}
\acro{ITS}{intelligent transport system}
\acro{IVIM}{infrastructure to vehicle information message}
\acro{LDM}{Local Dynamic Map}
\acro{LOS}{line-of-sight}
\acro{MAC}{medium access control}
\acro{MCM}{maneuver coordination message}
\acro{MIMO}{multiple input multiple output}
\acro{Moveover}{seamless mobility of vehicles over intersections}
\acro{MQTT}{message queuing telemetry transport}
\acro{NLOS}{non-line-of-sight}
\acro{NMEA}{National Marine Electronics Association}
\acro{OBU}{on-board unit}
\acro{OFDM}{orthogonal frequency-division multiplexing}
\acro{OScar}{Open Stack for Car}
\acro{PHY}{physical}
\acro{PVT}{Position Velocity Time}
\acro{ROS2}{Robot Operating System~2}
\acro{RSU}{road side unit}
\acro{TLM}{Traffic Light Maneuver}
\acro{V2N}{vehicle-to-network}
\acro{V2I}{vehicle-to-infrastructure} 
\acro{V2V}{vehicle-to-vehicle} 
\acro{V2X}{vehicle-to-everything} 
\acro{VAM}{vulnerable road user awareness message}
\acro{VESC}{Vedder Electronic Speed Controller}
\acro{WAVE}{wireless access in vehicular environment}
\end{acronym}

\begin{document}
\bstctlcite{IEEEexample:BSTcontrol}
%
\title{A Practical Implementation of Day-3 Cooperative Intersection with Automated Connected Mini-Cars
}

\author{\IEEEauthorblockN{
Lorenzo Farina\IEEEauthorrefmark{1}\IEEEauthorrefmark{2}, 
Vittorio Todisco\IEEEauthorrefmark{1}\IEEEauthorrefmark{2}, 
Federico Gavioli\IEEEauthorrefmark{3}, 
Salvatore Iandolo\IEEEauthorrefmark{4}, 
Francesco Moretti\IEEEauthorrefmark{3},\\
Giuseppe Perrone\IEEEauthorrefmark{6}, 
Matteo Piccoli\IEEEauthorrefmark{2}, 
Francesco Raviglione\IEEEauthorrefmark{5}, 
Marco Rapelli\IEEEauthorrefmark{6}, 
Antonio Solida\IEEEauthorrefmark{4},\\
Claudio Casetti\IEEEauthorrefmark{6}, 
Paolo Burgio\IEEEauthorrefmark{3}, 
Carlo Augusto Grazia\IEEEauthorrefmark{4}, 
and Alessandro Bazzi\IEEEauthorrefmark{1}\IEEEauthorrefmark{2}
}\\
\IEEEauthorblockA{\IEEEauthorrefmark{1}DEI, Universit\`a di Bologna, 40136 Bologna, Italy}
\IEEEauthorblockA{\IEEEauthorrefmark{2}National Laboratory of Wireless Communications (WiLab), CNIT, 40136 Bologna, Italy}
\IEEEauthorblockA{\IEEEauthorrefmark{3}FIM, University of Modena and Reggio Emilia, 41125 Modena, Italy}
\IEEEauthorblockA{\IEEEauthorrefmark{4}DIEF, University of Modena and Reggio Emilia, 41125 Modena, Italy}
\IEEEauthorblockA{\IEEEauthorrefmark{5}Dipartimento di Elettronica e Telecomunicazioni, Politecnico di Torino, 10129 Torino, Italy}
\IEEEauthorblockA{\IEEEauthorrefmark{6}Dipartimento di Automatica e Informatica, Politecnico di Torino, 10129 Torino, Italy}
}


\markboth{Journal of \LaTeX\ Class Files,~Vol.~14, No.~8, August~2015}%
{Shell \MakeLowercase{\textit{et al.}}: Bare Demo of IEEEtran.cls for IEEE Transactions on Magnetics Journals}
%





\IEEEtitleabstractindextext{%
\begin{abstract}
Cooperative driving enabled by connected and automated vehicles is expected to improve traffic efficiency and safety, particularly at intersections where traditional control mechanisms such as traffic lights introduce delays and unnecessary stops. Although cooperative intersection management algorithms have been widely studied, experimental demonstrations remain limited. This paper presents a real-time demonstration of cooperative intersection management using connected autonomous mini-cars. The testbed consists of multiple 1:10 scale vehicles equipped with autonomous driving capabilities and wireless communication modules that interact with a centralized controller responsible for scheduling their crossing of the intersection. Vehicles approaching the intersection exchange messages with the controller to set the appropriate mobility profile to traverse the intersection without stopping. The demonstration integrates autonomous driving, wireless communication, and cooperative control in a single experimental platform, providing a practical environment for validating cooperative intersection management concepts for future intelligent transportation systems.
\end{abstract}

\medskip
}

\maketitle

\IEEEdisplaynontitleabstractindextext

%
\IEEEpeerreviewmaketitle

\acresetall


\section{Introduction}

Recent advancements in sensing, automation, and wireless communication are enabling \acp{CAV} 
to exchange information with other vehicles and roadside infrastructure through \ac{V2X} technologies. These capabilities support cooperative driving strategies in which vehicles adapt their behavior based on shared information.
The development of \ac{C-ITS} services is often described through a progressive roadmap referred to as Day~1, Day~2, and Day~3. Day~1 applications focus on cooperative awareness through the periodic exchange of vehicle status information. Day~2 services extend this paradigm through collective perception, where vehicles share information about objects detected by their onboard sensors. Day~3 introduces cooperative maneuver coordination, where vehicles exchange information about their intended actions and coordinate their trajectories to perform complex maneuvers. In Europe, such coordination is expected to be supported by Maneuver Coordination Messages (MCM), currently under standardization within ETSI.

In the context of Day-3 services, 
coordinated maneuvers will be a cornerstone to improve road efficiency, especially in cities, where cooperation at intersection can increase road capacity significantly\cite{tachet2016revisiting}. Road intersections therefore represent one of the most relevant yet challenging scenarios for cooperative driving. At these locations, multiple traffic flows interact, and traditional control mechanisms such as traffic lights or right-of-way rules are designed primarily to guarantee safety rather than to optimize traffic efficiency. As a result, vehicles are often required to stop or wait even when no immediate conflicts are present.

Recently, several studies have  investigated cooperative approaches to intersection management, where \acp{CAV} exchange information about their planned trajectories and adapt their motion accordingly \cite{11202555}. Proposed solutions include reservation-based schemes~\cite{wang2021digital}, trajectory optimization methods~\cite{zhang2019decentralized}, and centralized coordination strategies~\cite{hult2018energy}. However, most of these studies rely on simulations or analytical models, while experimental validation remains comparatively limited due to the complexity of integrating autonomous driving, communication technologies, and coordination algorithms in real systems. 
Although some advanced field trials have been reported, for example in \cite{8790807,9745747}, these experiments typically involve only two or three vehicles and consider a single crossing event under highly controlled conditions.




To mitigate the cost and safety challenges associated with real-world testing, experimental platforms based on scaled autonomous vehicles provide a practical way to investigate cooperative driving concepts. Such platforms, which have been primarily used to study autonomous driving functionalities \cite{10817784}, allow researchers to integrate sensing, communication, and control components within a manageable and reproducible 
setup.


\begin{figure*}[t]
\centering
\includegraphics[width=1.7\columnwidth, draft=false]{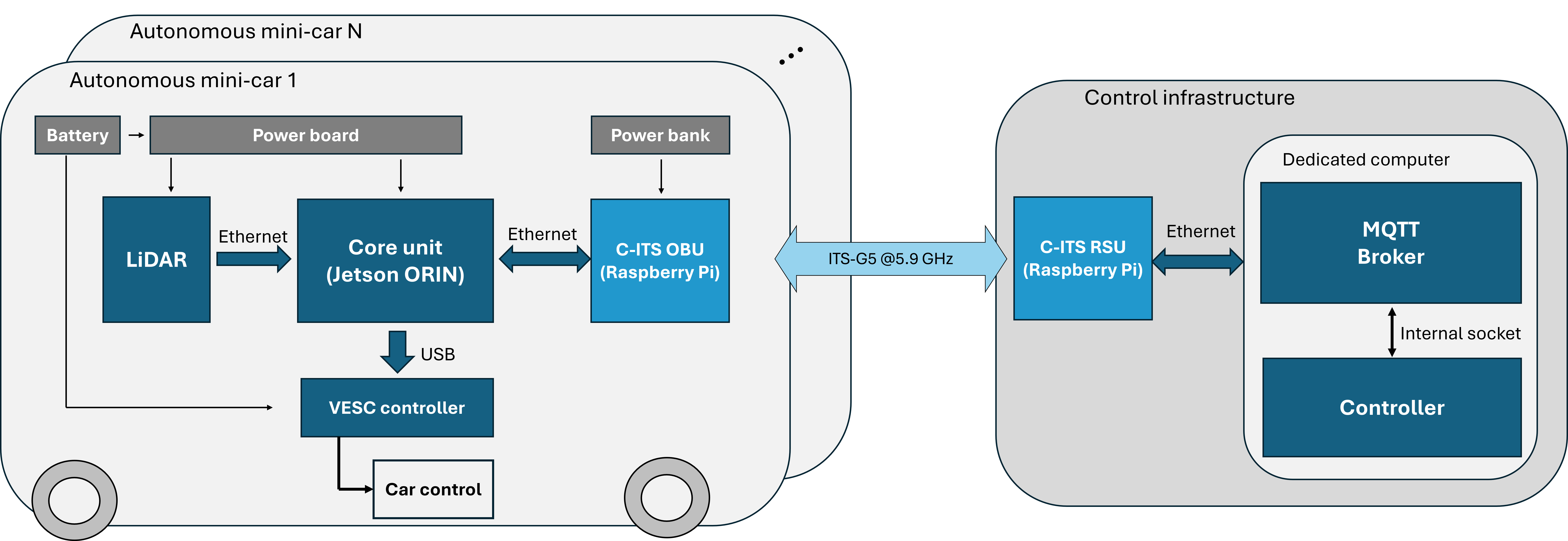}
\caption{Testbed architecture.} \label{fig:architecture}
\end{figure*}

In this paper, we present an experimental demonstration of cooperative intersection management using connected autonomous mini-cars. The demonstration involves multiple 1:10 scale \acp{CAV} operating on a figure-eight track and interacting with a centralized controller. Communication between vehicles and the controller is implemented using \ac{MQTT} over ITS-G5 in the ITS band at 5.9~GHz. As a representative example of the capabilities of the platform, the system implements the coordination algorithm described in \cite{farina2026v2nbasedalgorithmcommunicationprotocol}, which is scalable and lightweight from a communication perspective.

The objective of this work is 
to demonstrate the integration of autonomous driving, wireless communication, and centralized scheduling within a single experimental platform. The resulting testbed provides a practical environment for studying cooperative intersection management and validating communication-assisted coordination mechanisms in a controlled setting.


\section{Testbed Architecture Overview}
\label{sec:architecture}

%
The system architecture consists of three main components: the autonomous mini-cars, the wireless communication subsystem, and the control infrastructure. A block scheme is provided in Fig.~\ref{fig:architecture}.


\subsection{Autonomous Mini-Cars}

The autonomous mini-cars are based on the Roboracer platform \cite{okelly2019f110opensourceautonomouscyberphysical} (formerly F1TENTH), which is an open-source framework widely used for autonomous driving research. The platform provides modular hardware and open source software components that can be adapted to a variety of experimental scenarios.

From a hardware perspective, each vehicle uses a commercially available 1:10 scale radio-controlled chassis equipped with a custom deck hosting the computing, sensing, and power subsystems. A \ac{VESC} controls both the brushless DC motor (speed) and the steering servomotor, providing motor abstraction and sensor feedback through an integrated BMI160 \ac{IMU}. Environmental perception is primarily provided by a Hokuyo UST-10LX 2D LiDAR sensor. 

The central computing unit is an NVIDIA Jetson Orin platform (either the 8-core NX or the 6-core Nano Super), which provides both CPU and GPU resources for running the autonomous driving stack.

The software architecture is built on \ac{ROS2} \cite{ros2}, leveraging its publish--subscribe communication model to enable a modular and extensible system. The main software components include sensor and \ac{VESC} drivers, a pure pursuit trajectory-following controller \cite{Wallace1985FirstRI}, and a custom localization module combining LiDAR and \ac{IMU} measurements. Future work will integrate an open-source localization component based on Monte Carlo particle filtering \cite{Fox2001}.

\subsection{Wireless Communication Subsystem
}

The cooperative intersection management system requires vehicles to exchange information with the control infrastructure. This communication is implemented through short-range vehicular wireless communication operating in the ITS band at 5.9~GHz.

In this work, we employ a compact open-source communication module previously developed to support IEEE~802.11p and ETSI ITS-G5 operation; a detailed description of the hardware architecture and its configuration can be found in \cite{11174862}. 
The core of the unit is a Raspberry Pi~5 Model~B coupled with a MikroTik R11e-5HnD wireless card based on the Atheros AR9580 chipset. Although originally designed for IEEE~802.11a/n operation, this chipset can be patched to support communication in the ITS band \cite{iandolo2025odu}.

The platform is capable of exchanging ETSI-compliant messages through the open-source OScar stack \cite{OScar_paper_2024,farina2026opensourcebasedetsicompliant}. However, this functionality is not used in the cooperative intersection management application presented in this paper, since maneuver coordination messages are still under standardization. Instead, vehicles and controller exchange application-level messages using the \ac{MQTT} protocol over TCP/IP.

Each mini-car is equipped with one of these units acting as an \ac{OBU}. An additional unit is deployed at the controller node and operates as a \ac{RSU}, enabling wireless communication between the vehicles and the centralized controller.

\subsection{Control Infrastructure}

The control infrastructure runs on a centralized processing node, implemented on a standard laptop. The node hosts two main software components.

\begin{itemize}
    \item The controller, which executes the scheduling algorithm responsible for coordinating the mini-cars and preventing collisions at the intersection. The algorithm is implemented in Python and its core logic is described in Section~\ref{sec:moveover}.

    \item An \ac{MQTT} broker used to support communication between the vehicles and the controller. The broker manages a set of topics to route messages: one shared topic for messages sent by the vehicles to the controller, and one dedicated topic per vehicle through which the controller sends coordination commands. The messaging service is implemented using the open-source \texttt{Eclipse Mosquitto} daemon.
\end{itemize}

\section{Algorithm for the Intersection Management}
\label{sec:moveover}

\begin{figure}[t]
\centering
\includegraphics[width=0.99\columnwidth]{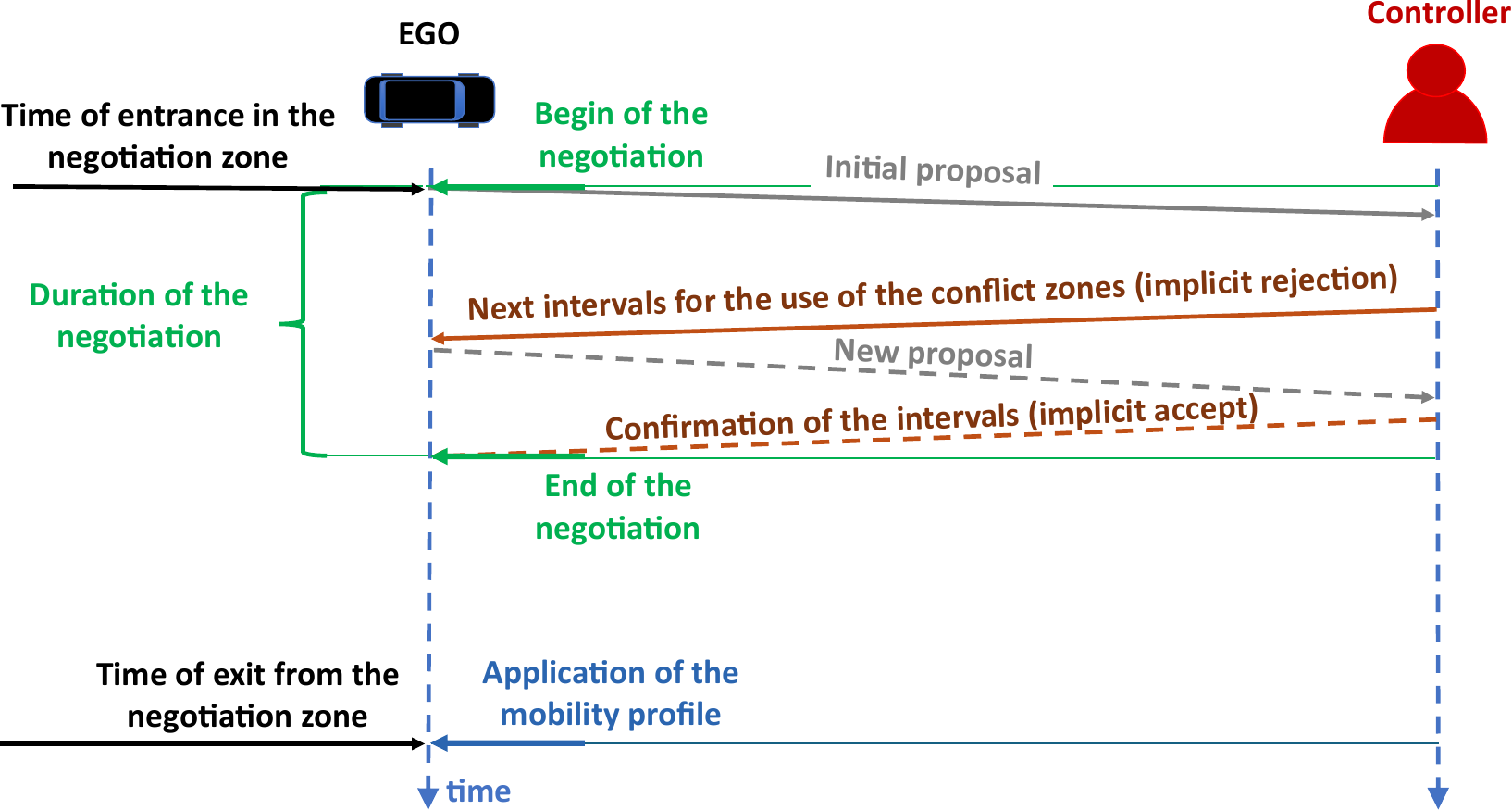}
\caption{Temporal sequence of the negotiation procedure between a vehicle and the controller. In this example, the initial proposal is rejected and the second proposal is accepted.}
\label{fig:exchange}
\end{figure}

The cooperative intersection management demonstrated in this work relies on a scheduling algorithm, called \ac{Moveover}, and the related communication protocol.
A detailed description of the algorithm can be found in \cite{farina2026v2nbasedalgorithmcommunicationprotocol}; here, we summarise the key aspects of the solution.

\medskip
The execution of \ac{Moveover} is distributed between the individual \acp{CAV} and the centralized controller. The goal of the algorithm is to determine, for each vehicle, a mobility~profile\footnote{For conciseness, we use the term mobility profile to denote a vehicle's trajectory together with its speed profile.} that allows it to traverse the intersection safely and without stopping, except in situations of traffic congestion.


Only one vehicle at a time can negotiate its mobility profile with the controller, and the \ac{CAV} performing this operation is referred to as the EGO. Once scheduled, the vehicle follows the assigned mobility profile during normal operation. 

If abnormal conditions occur, such as communication failures or unexpected traffic situations, the system switches to the backup mode described in Section~\ref{sec:backup}.

\subsection{Zones}\label{sec:zones}

\ac{Moveover} models the intersection using two distinct areas, called the \emph{negotiation zone} and the \emph{conflict zone}. This abstraction allows the algorithm to be applied to intersections with different geometries.

\begin{itemize}
    \item Negotiation zone: the area where the EGO and the controller negotiate the mobility profile. One negotiation zone is defined for each incoming lane, and the vehicle must complete the negotiation before leaving this area; otherwise, the backup mode is triggered.
    
    \item Conflict zone: a road portion inside the intersection that can be occupied by only one vehicle at a time. The controller prevents simultaneous access to the same conflict zone by enforcing strictly non-overlapping temporal reservations.
\end{itemize}

Upon entering the negotiation zone, a \ac{CAV} initiates the negotiation procedure, but it effectively becomes the EGO only when previously entered \acp{CAV} have completed the negotiation. Vehicles are therefore scheduled according to a \ac{FIFS} policy, based on their arrival order.

\subsection{Communication Protocol}\label{sec:protocol}

The communication protocol consists of a negotiation loop between the EGO and the controller to determine a collision-free mobility profile.
The negotiation procedure is illustrated in Fig.~\ref{fig:exchange}, which shows an example where the controller requests the EGO to delay its entry into the conflict zones.

\smallskip

\subsubsection{Proposal by the EGO}
Upon entering the negotiation zone, the EGO calculates an initial mobility profile based strictly on its own physical capabilities and constraints (e.g., maximum road speed, safe turning speeds, and comfort-based acceleration/deceleration limits). This profile is then submitted to the controller.

\subsubsection{Validation by the controller}
The controller cross-references the EGO's proposal against already scheduled vehicles across three stages:
\begin{itemize}
    \item Check for conflicts before the intersection, ensuring a minimum safe following distance behind the vehicle ahead of the EGO in the same lane;
    \item Check for conflicts within the intersection, verifying that the EGO occupies the conflict zones in time intervals that do not overlap with existing reservations;
    \item Check for conflicts after the intersection, ensuring that the EGO's exit speed and distance are safely compatible with the vehicle exiting ahead of it.
\end{itemize}

\subsubsection{Response from the controller}

The controller replies with permissible entry and exit time bounds for each of the conflict zones to be traversed. Acceptance or rejection is implicit in the returned message: if the EGO's proposed mobility profile complies with the specified time windows, the EGO infers that its proposal has been accepted and the controller records the reservation. Otherwise, the EGO interprets the response as a rejection and must recompute its mobility profile using the newly provided bounds as constraints. The new bounds proposed by the controller are always later than those originally proposed by the EGO.

\begin{figure*}[t]
\centering
\includegraphics[width=1.75\columnwidth]{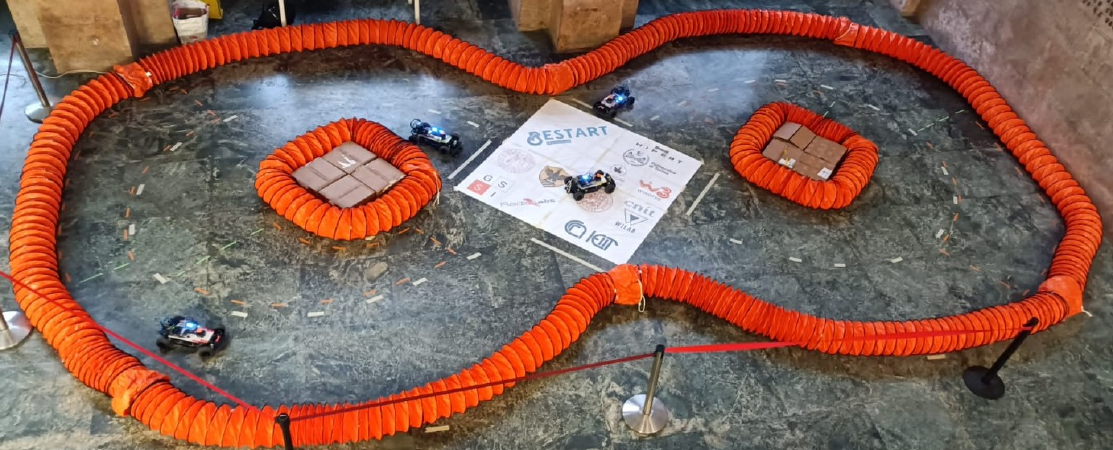}
\caption{Implemented environment for the evaluation of cooperative intersection management.} 
\label{fig:scenario}
\end{figure*}


\subsubsection{Additional exchanges}

Upon rejection, the EGO uses the controller's availability windows to compute and submit a new, delayed mobility profile. This back-and-forth negotiation may iterate multiple times until an agreement is reached. In practice, the number of exchanges is typically small, since each response from the controller can only delay the entry time into the first conflict zone.


\subsection{Backup Mode}\label{sec:backup}

The constraint that the controller can only introduce delays guarantees that the negotiation process converges to a deterministic conclusion. However, communication delays may prevent the exchange from completing before the EGO exits the negotiation zone. If the EGO cannot finalize its mobility profile before leaving the negotiation zone, or if the required crossing speed falls below a minimum threshold, the system switches to a backup mode. In this mode cooperative scheduling is suspended and vehicles revert to autonomous driving using conventional right-of-way priorities.

\section{Testbed Scenario and Experimental Setup}
\label{sec:demo}

To validate the proposed system, we designed a physical testbed that reproduces a cooperative intersection scenario in a manageable environment. 
The setup consists of multiple autonomous mini-cars moving along a closed track and repeatedly approaching a shared intersection. The objective is to evaluate the practical viability of \ac{Moveover} by demonstrating how vehicles can coordinate with the controller to cross the intersection without stopping.
A summary of the key experimental parameters, including track dimensions and vehicle specifications, is provided in Table~\ref{tab:params}.

\begin{figure}[t]
\centering
\includegraphics[width=0.925\columnwidth]{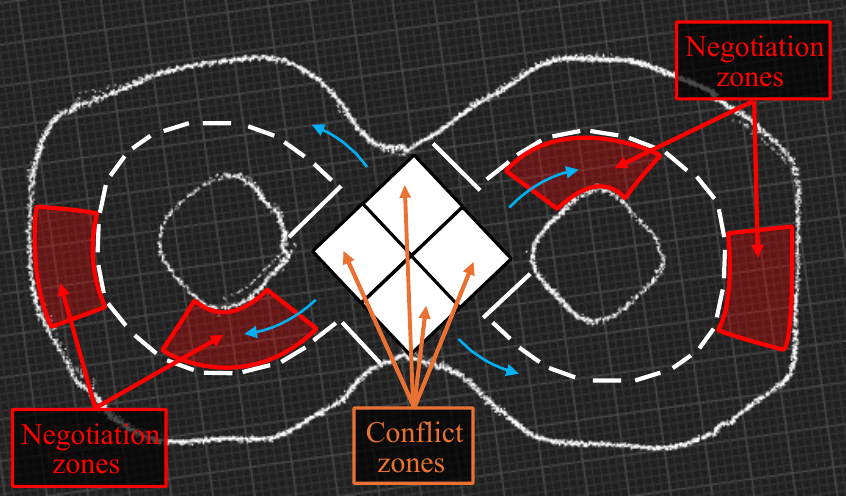}
\caption{Screenshot of the map with highlighted negotiation and conflict zones.}
\label{fig:map}
\end{figure}

    	\begin{table}[t]
	\caption{Characteristics of the experimental setup.}
    \label{tab:params}
		\centering 
        \small
	\begin{tabular}{p{5cm}p{0.9cm}}
\hline \hline
\textbf{Parameter} & \textbf{Value} \\ \hline 
\textbf{\textit{Track}}\\
Length & 11~m \\ 
Width & 8~m \\ 
Minimum length & 8.20~m \\ 
Minimum width & 4.20~m \\ 
Conflict zone side & 0.9~m \\ 
Negotiation zone length & 2~m \\
 \hline
\textbf{\textit{Mini-cars}}\\
Length & 0.5~m \\ 
Width & 0.3~m \\ 
Maximum speed & 1~m/s \\
\hline \hline
\end{tabular}
	\end{table}

\begin{figure*}[t]
\centering
\subfloat[A and B approach the intersection]{%
\includegraphics[width=0.32\textwidth,trim=3cm 0.6cm 0 0,clip]{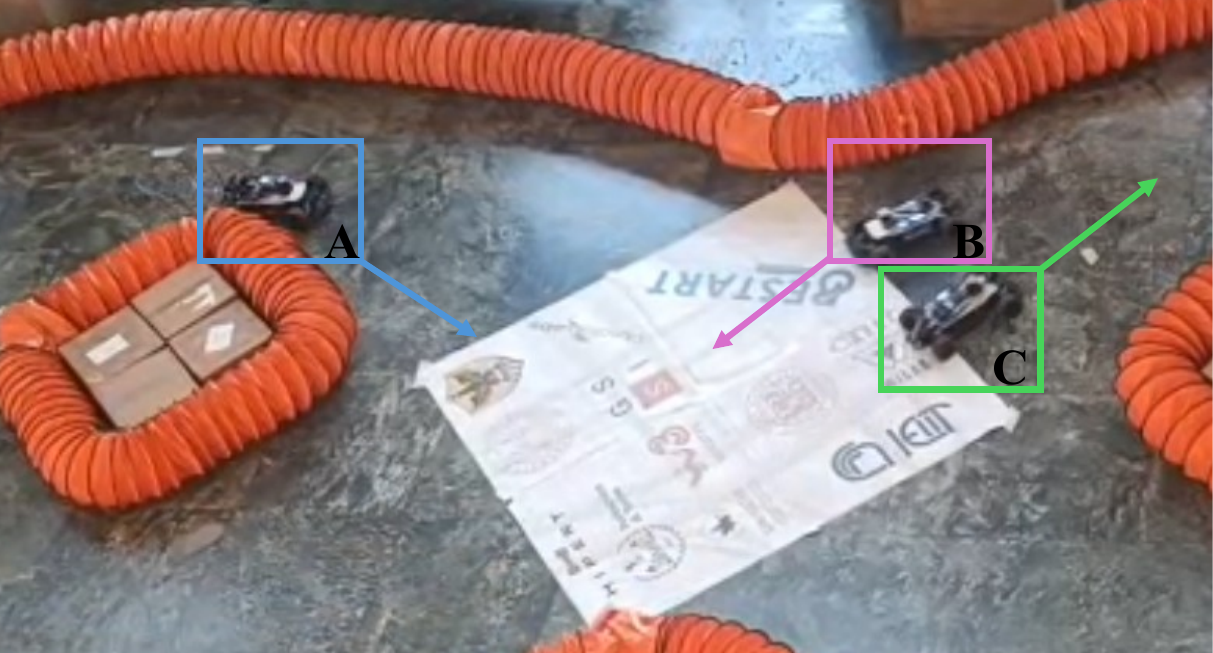}}
\hfill
\subfloat[B is scheduled to traverse first]{%
\includegraphics[width=0.32\textwidth,trim=3cm 0.6cm 0 0,clip]{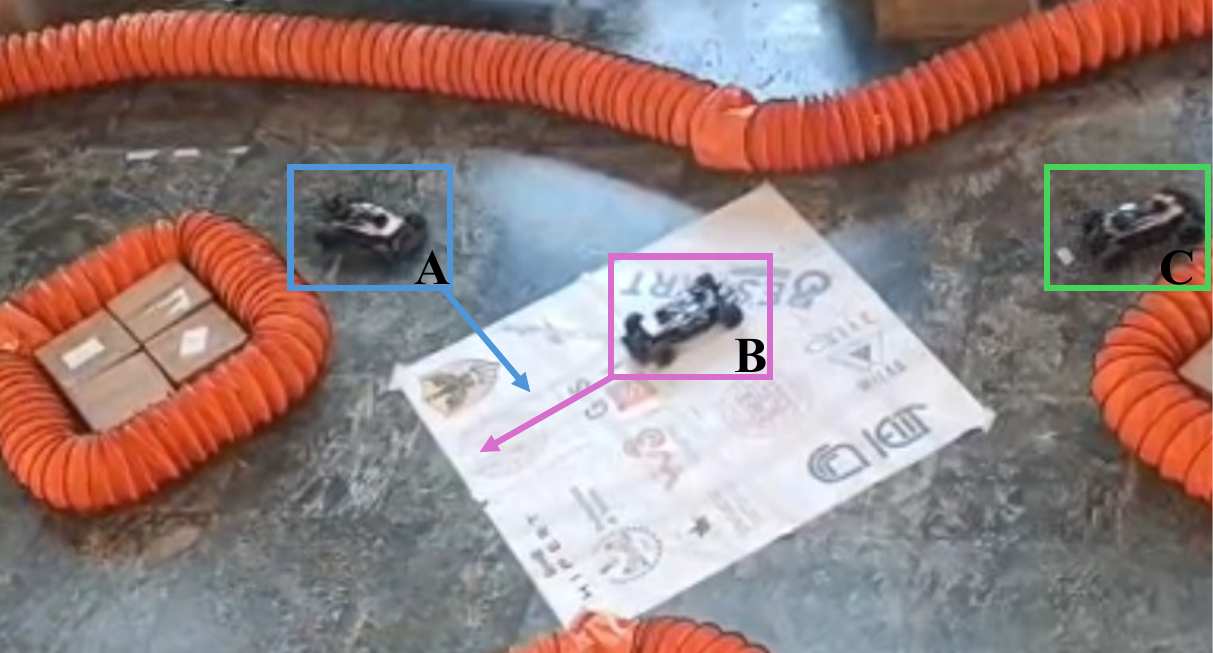}}
\hfill
\subfloat[A crosses the intersection as B leaves]{%
\includegraphics[width=0.32\textwidth,trim=3cm 0.6cm 0 0,clip]{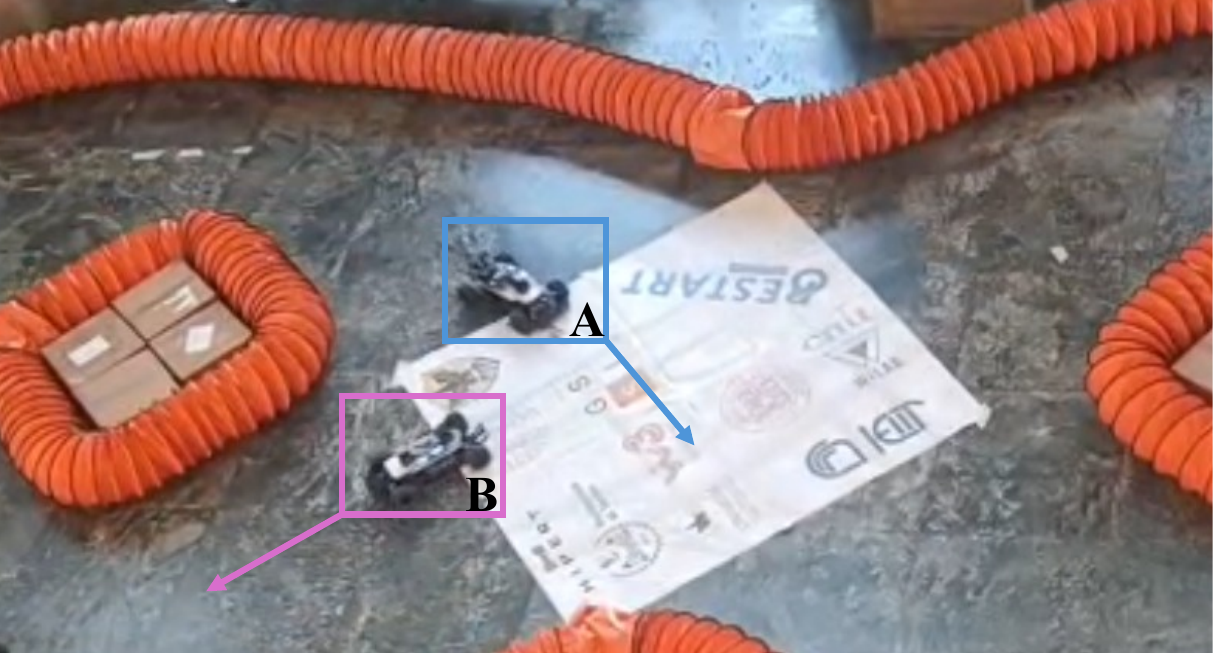}}
\caption{Sequence of images from the cooperative intersection demonstration. Multiple mini-cars approach the intersection and traverse it without stopping under the coordination of the controller. The full video is available at \cite{Todisco2026}.}
\label{fig:demo_sequence}
\end{figure*}

\subsection{Track Layout}

The demonstration is conducted on a figure-eight track, shown in Fig.~\ref{fig:scenario}. The track has a length of approximately 11~m and a width of 8~m, which can be reduced to a minimum of approximately 8.20~m by 4.20~m with lower speed of the mini-cars.

%
Flexible orange tubes are used to delimit the track and create two loops that intersect at a central crossing point. The intersection area is subdivided into four square regions corresponding to the conflict zones, each with a side length of 0.9~m.
%
Square prints are used to visualize the conflict zones, although they are not used by the mini-cars or the controller. Similarly, tape markings on the floor indicate the lanes and the negotiation zones, which are defined for each lane approaching the intersection. Each negotiation zone has a length of 2~m.



\subsection{Demonstration Workflow}

During the demonstration, the mini-cars continuously circulate along the figure-eight track and repeatedly encounter the intersection. Each vehicle follows a predefined trajectory while the onboard autonomous driving system ensures accurate path tracking. In our experiments, four mini-cars were used, each measuring 0.5~m in length and 0.3~m in width, operating at a maximum speed of 1~m/s. To generate diverse yet repeatable interaction scenarios, two vehicles always proceed straight through the intersection, one always turns right, and one always turns left.

Accurate path tracking requires continuous self-localization. For this purpose, each mini-car is equipped with a map of the track generated prior to the experiments by manually driving a vehicle along the track while recording LiDAR data. The resulting point cloud is refined offline and uploaded to all vehicles. An example of the resulting map for the scenario in Fig.~\ref{fig:scenario} is shown in Fig.~\ref{fig:map}. The same map is also used to design the reference trajectories followed by the mini-cars, using proprietary \ac{CAD} software.

Once the map and reference trajectories are loaded, the mini-cars can navigate the track autonomously. When entering a negotiation zone, a mini-car transmits an initial proposal to the controller, initiating the negotiation procedure described in Section~\ref{sec:protocol}.

In the current implementation, if a vehicle triggers the backup mode, centralized scheduling is suspended and all mini-cars stop. During our experiments, this situation occurred only rarely during long test runs, typically due to battery depletion in one of the vehicles. Alternative implementations of the backup mode are left for future work.


Using these specific settings, the system sustained autonomous, collision-free operation for more than one hour, halting only due to onboard battery exhaustion. The four 1:10 \acp{CAV} operated simultaneously with full localization, perception, and control loops, coordinated by the central infrastructure. The system maintained stable connectivity, consistent timing in command execution, and precise synchronization between the vehicles and the intersection controller. These experiments confirm that the integrated platform achieves a high level of reliability and is suitable for extended experimental studies.

Fig.~\ref{fig:demo_sequence} shows a sequence of instances from the demonstration, where two mini-cars approach the intersection and traverse it without stopping under the coordination of the controller. A video of the demonstration is available at \url{https://www.youtube.com/watch?v=Mt51gEt45E0}.

\section{Conclusion}\label{sec:conclusion}

This work demonstrated, through a scaled experimental testbed, the feasibility of cooperative maneuvers among connected and autonomous vehicles at non-stop non signalized intersections. The demonstration integrates perception, control, and communication components within a unified experimental platform.
The results confirm the practical applicability of the intersection coordination algorithm used in the system and highlight the suitability of the proposed platform as a research and validation environment for cooperative intersection management. More broadly, the testbed provides a flexible framework for experimental studies in the field of cooperative, connected, and automated mobility.


\section*{Acknowledgment} This work was partially supported by the European Union under the Italian National Recovery and Resilience Plan (NRRP) of NextGenerationEU, partnership on ``Telecommunications of the Future'' (PE00000001 - RESTART), project MoVeOver, and national research centre on mobility (CN00000023 - MOST).



\bibliographystyle{IEEEtran}  
\bibliography{biblio}

\end{document}